\documentclass[12pt]{article}
\usepackage{amssymb}
\usepackage{amsfonts}
\usepackage{latexsym}

\usepackage{amsmath}
\usepackage{graphicx}

\def\beq{\begin{eqnarray}}
\def\eeq{\end{eqnarray}}
\def\ln{\,\mbox{ln}\,}

\def\al{\alpha}
\def\be{\beta}

\def\ga{\gamma}
\def\de{\delta}
\def\vp{\varepsilon}
\def\ep{\epsilon}
\def\ze{\zeta}

\def\la{\lambda}
\def\na{\nabla}

\def\si{\sigma}

\def\ph{\varphi}

\def\Ga{\Gamma}
\def\De{\Delta}

\begin{document}
\begin{center}

{\large\sc 
Local conformal symmetry and its fate at quantum level
\footnote{Talk presented at the Fifth International Conference 
on Mathematical Methods in Physics, 24 - 28, April 2006,
Rio de Janeiro, Brazil. To be published by PoS/JHEP.}}
\vskip 6mm

{\bf Ilya L. Shapiro}
 \footnote{
Also at Tomsk State Pedagogical University, Tomsk, Russia; \
electronic address: shapiro@fisica.ufjf.br}

\vskip 2mm

{\small\sl Departamento de F\'{\i}sica -- ICE, 
Universidade Federal de Juiz de Fora, MG, Brazil}

\end{center}
\vskip 6mm

\begin{quotation}

\noindent
{\large \sl Abstract.}
The purpose of this article is to present a short review
of local conformal symmetry in curved $4d$ space-time. 
Furthermore we discuss the conformal anomaly and 
anomaly-induced effective actions. Despite the conformal 
symmetry is always broken at quantum level, it may be 
a basis of useful and interesting approximations for
investigating quantum corrections. 
\end{quotation}
\vskip 8mm

\section{Introduction}

The local conformal symmetry of matter fields in curved space
and proper gravity always attracted a great interest. It is 
not our aim to list the main publications on the 
subject nor the main lines of research related to it. 
So, let us start by mentioning a recent review \cite{faraoni} 
where one can find some relevant references to start. In our 
article we shall concentrate on those aspects of the conformal 
theory in four dimensions, which are relevant for the 
applications, especially to cosmology. We shall pay special 
attention to the quantum theory and discuss conformal anomaly 
and anomaly-induced effective action of gravity. Many other 
issues will be left aside, some of them may be eventually 
considered in the extended version of this review article. 
 
In order to understand the reason to introduce a local 
conformal symmetry, let us start from a very simple example 
discussed in \cite{anju}. Consider a massive scalar field 
in curved space-time. The minimal action has the form 
\beq
S=\frac12\,\int\sqrt{-g}\Big(
g^{\mu\nu}\na_\mu\ph\na_\nu\ph + m^2\ph^2 + \xi R\ph^2\Big)\,,
\label{scal}
\eeq
where $\xi=0$. However, the minimal theory is inconsistent at 
quantum level if we introduce interactions with other fields 
or scalar self-interaction. In principle, one has to keep the 
non-minimal parameter $\xi$ arbitrary to provide the 
multiplicative renormalizability of the theory. At the same 
time the value $\xi=1/6$ is very special, for in the massless 
case $m=0$ it corresponds to the local conformal symmetry 
\beq
g_{\mu\nu} \to g_{\mu\nu}^\prime = g_{\mu\nu} e^{2\si(x)}
\,\quad
\ph \to \ph^\prime = \ph e^{-\si(x)}\,.
\label{trans}
\eeq
Now, let us consider the massless limit of the theory from 
another point of view. Basing on fundamental principles of 
quantum theory, one is expecting to meet correspondence 
between the field and particle description of the matter. 
It is well known that, for the classical particles, the 
massless limit corresponds to the vanishing trace of the 
energy-momentum tensor 
\beq
T^\mu_\mu=-\frac{2}{\sqrt{-g}}\,
g_{\mu\nu}\frac{\de S}{\de g_{\mu\nu}}=0\,. 
\label{loc}
\eeq
However, this identity can be achieved, in the field 
description (\ref{scal}), only for the conformal value 
$\xi=1/6$ of the non-minimal parameter (of course, the 
relation (\ref{loc}) holds only on the mass shell for 
the scalar field or for the corresponding particles). 
Therefore, only conformal theory can provide a correct 
particle-field correspondence in the massless limit. 
One can see that the conformal value \ $\xi=1/6$ \ does 
provide certain advantage at this level. The next question 
is whether it is possible to maintain the conformal value 
of \ $\xi$ \ and, in general, local conformal symmetry, at 
the quantum level, when the loop corrections are taken into 
account. This issue is the main subject of the present review. 

The paper is organized as follows. 
In the next section we shall list known conformal 
theories in $n=4$ and after that consider the quantum 
theory, where the local conformal invariance always 
breaks down. Section 3 is devoted to the brief 
description of the anomaly. We discuss the main 
difference between global and local conformal symmetries 
at quantum level. In section 4 we present, in more details 
than usual, the derivation of the anomaly-induced effective
action of vacuum. In section 5 we come back to anomaly and 
describe its ambiguities in relation to the effective 
action. Section 6 contains a brief description of the 
situation in conformal quantum gravity. Finally, in 
section 7 we draw our conclusions. 

\section{Particular examples of conformal theories}
 
Consider a general metric-scalar theory \cite{conf}
\beq
S = \int d^4x\sqrt{-g}\; \left\{ 
A(\phi)\,\left(\na\phi\right)^2 + B(\phi)R  + C(\phi) 
\right\}\,,                                                    
\label{action}
\eeq
and perform the local conformal transformation of $g_{\mu\nu}$ 
plus an arbitrary scalar reparametrizations (thus, generalizing
the eq. (\ref{trans})). In order to make things simpler, we 
start from the action without kinetic term for the scalar 
field \cite{ohanlon} and transform it to (\ref{action})
\beq
S = \int d^4x \sqrt{g'} 
\,\left\{ R'\Phi+ V(\Phi)\right\}\,,\quad                                                    
\label{new action}
g'_{\mu\nu}=g_{\mu\nu}e^{2\sigma(\phi)}
\,,\quad \Phi=\Phi(\phi)\,.
\eeq
Simple calculation leads to the relation
\beq
A(\phi)=6e^{2\sigma(\phi)}[\Phi\sigma_1+\Phi_1]\sigma_1
\,,\quad
B(\phi)=\Phi(\phi) e^{2\sigma}\,,                     
\label{AB}
\eeq
where \ $B_1 = {dB}/{d\phi}$ \ etc. One can see that the 
absence of the kinetic term in the action 
(\ref{new action}) does not mean that this field is not 
dynamical. The dynamics of the scalar field is due to 
the interaction with the scalar mode of the metric. For
instance, the free minimal scalar field plus General 
Relativity (GR) is the particular case of the action 
(\ref{action}) and is conformally equivalent to 
(\ref{new action}). 

The conformal symmetry of the action corresponds to 
pure GR, $\Phi = const$. Then
\beq
A=\frac{3B_1}{2B^2}\,\,,\quad C=\la B^2\,,
\quad\mbox{where}\quad
B=B(\phi)\,,\,\,\,C=C(\phi)\,,\,\,\,
B_1=\frac{dB}{d\phi}\,.
\label{condi}
\eeq
The well-known particular case of the theory satisfying the 
constraints (\ref{condi}) is (\ref{scal}) with \ $m=0$, 
$\xi=1/6$ \ and with an additional self-interaction term. 
One can rewrite it in the form 
 \beq
 S = \frac12\,\int d^4x \sqrt{-g}\; \Big(\,-
 \phi\,\De_2\,\phi + \frac{\la}{12}\,\phi^4 \,\Big)
\,,\quad
 \mbox{{where}}\quad
 \De_2\,=\,\Box - R/6\,.
\label{usual}
 \eeq
All models which satisfy (\ref{condi}) are linked by conformal 
transformation of the metric plus scalar reparametrization 
\cite{conf}.

Other conventional examples of conformal fields include 
massless spinor and vector 
\beq
S_{1/2}\,=\,\frac{i}{2}\,\int d^4x\sqrt{g}\,\left\{\,
\bar{\psi}\,\ga^\mu\,\na_{\mu}\psi 
\,-\, \na_{\mu}\bar{\psi}\,\,\ga^\mu \psi
\,\right\}\,,
\\
S_{1} \,=\, -\, \frac14\,\int d^4 x\sqrt{g}\,
 F_{\mu\nu}F^{\mu\nu}\,,
\label{spi vec}
\eeq
with the transformation rules 
\beq
g_{\mu\nu} \to g_{\mu\nu}^\prime = g_{\mu\nu}\,e^{2\si}
\,,\quad
A_\mu \to A_\mu^\prime = A_\mu
\,,\quad
\psi \to \psi^\prime=\psi\,e^{-3\si/2}
\,,\quad
{\bar \psi} \to {\bar \psi}^\prime={\bar \psi}\,e^{-3\si/2}\,.
\eeq
Let us notice that the difference between conformal 
weight and dimension for the vector field is due to
the vector field definition in curved space-time\footnote{I am 
grateful to Joan Sol\`a, who called my attention to this point.}
\beq
A_\mu = A_b\,e^b_\mu
\,,\quad
e^b_\mu\,e^a_\nu\,\eta_{ab}=g_{\mu\nu}\,,
\quad
e^b_\mu\,e^a_\nu\,g^{\mu\nu}=\eta^{ab}\,.
\label{vec}
\eeq
The importance of this observation is that it shows a direct 
relation between local and global conformal symmetries. 
The generalization to the non-Abelian case is straightforward. 

The interactions between usual vectors, scalars and fermions 
are always conformal if the corresponding coupling constants 
have zero mass dimension. Hence gauge, Yukawa and \ $\phi^4$
\ interaction terms are conformal while  \ $\phi^3$ \ is not. 

The last conventional example is the conformal (Weyl) 
gravity, which includes only metric field
\beq
S_W\,=\,\int d^4x\sqrt{g}\,C^2(4)\,,
\quad 
C^2(n)\,=\,
R^2_{\mu\nu\al\be} - \frac{4}{n-2}\,R^2_{\mu\nu}
+\frac{1}{(n-1)(n-2)}\,R^2\,,
\label{Weyl}
\eeq
where $n$ is the space-time dimension. The main difference
between the Weyl gravity model (\ref{Weyl}) and GR is that 
the former does not produce the correct Newton limit. This 
is of course natural because the coupling constant in this 
theory is dimensionless and therefore an additional 
mechanism is needed in order to produce a dimensional 
parameter such as Newton constant. The most natural option 
is to consider the Weyl gravity and the conformal scalar 
field together. In this case the quantum effects lead to 
the complicated effective potential for the scalar field. 
This may produce a dimensional transmutation and eventually 
lead to induced GR with induced values of both Newton and 
cosmological constants. A general review on the induced 
gravity approach can be found in \cite{Adler}. There are 
several possible mechanism of how this method can be 
applied to the initially conformal theory 
\cite{Buch-Weyl,induce,brv}. We will not discuss this aspect 
of the conformal theory in what follows, because this review 
is of a short kind. Instead, we shall concentrate on a more 
basic phenomenon (conformal anomaly) in the next section. 

The main difference between the conformal scalar, fermion 
and vector 
cases presented above and the last example of Weyl gravity 
is that it is a fourth derivative theory while the matter 
fields cases are all described by lower derivative theories. 
However one can construct also examples of conformal 
higher derivative scalars and fermions (and perhaps vectors,
despite this has not been done yet) which possess the local 
conformal invariance.

Let us start with scalars and consider two alternative 
different models. The fourth derivative scalar of the first 
kind has an action \cite{Paneitz,rei}
\beq
S_4&=&\int d^4x\sqrt{g}\,\,\ph\,\De_4\,\ph\,,
\label{reigert}
\\
\mbox{where} \qquad
\De_4 &=& \Box^2 + 2R^{\mu\nu}\na_\mu\na_\nu
- \frac23\,R\Box +\frac13\,R_{;\mu}\,\na^\mu\,.
\label{rei op}
\eeq
The conformal transformation law for this scalar is
 \ $\varphi \to \varphi^\prime$. 
The importance of the model (\ref{reigert}) is based on 
its use for integrating conformal anomaly. We shall discuss
this point in the next section. 

The fourth derivative scalar of the second kind can be 
presented, up to reparametrization of scalar
 \ $\chi=\chi(\phi)$, \ in the form \cite{ai}
\beq
\int d^{4}x\sqrt{-g}\;\left\{ 
\left( \phi^{-1}\,\Delta _{2}\,\phi \right)^2 
- a \phi \De_2 \phi - b\,\phi^4  + c\,C^{2} + d\,E \,\right\},
\label{ai}
\eeq
where \ $a,b,c,d$ \ are some constants. 
This model is a direct higher derivative generalization 
of the usual conformal scalar theory (\ref{usual}) and 
the transformation laws for metric and scalar are of 
course identical. The complete form of the 
parametrization-invariant higher derivative action, 
similar to (\ref{action}) with the constraints (\ref{AB})
satisfied, has been constructed in \cite{acacio}. One can
notice that the above two theories represent very particular 
cases of the general fourth derivative metric-scalar 
model formulated in \cite{eli}. This general model 
involves 12 arbitrary functions of the scalar (in fact
11, because one may be always included into scalar 
parametrization), while both models presented above 
have no such functions. 

Let us remark that both fourth derivative scalar models
(\ref{reigert}) and (\ref{ai}) can be generalized to an 
arbitrary dimension \ $n\neq 2$. For the case of 
(\ref{reigert}) this task has been completed in 
\cite{acacio} and for the case of (\ref{ai}) the 
procedure is obvious due to the known prescription 
for the usual conformal scalar (\ref{usual}).

The next example is a third derivative 
spinor field. In this case, again, the conformal 
invariance is provided by introducing the higher
derivatives, changing the transformation law for the 
field and adjusting the parameters of the higher
order differential operator  \cite{FrTs85,GBPISh97}.
The action of the model is
\beq
S_3\,=\,\frac{i}{2}\,\int d^4x\sqrt{-g}\,\left\{\,
 \bar{\psi}\ga^{\mu}{\cal D}_{\mu}\psi
 -{\cal D}_{\mu}\bar{\psi}\,\ga^{\mu}\psi\,\right\}\,,
\label{FT}
\eeq
where the self-adjoint third order operator has the form
\beq
{\cal D}_\mu = \na _{\mu}\Box + R_{\mu\nu}\na^{\nu}
- \frac{5}{12}\,R\na_{\mu} -\frac{1}{12}\,(\na_{\mu}R)\,.
\label{D}
\eeq
The transformation law for the spinor \ $\psi$ \ is
$$
\psi \to \psi^\prime=\psi\,e^{-\si/2}
\,,\qquad
{\bar \psi} \to {\bar \psi}^\prime={\bar \psi}\,e^{-\si/2}
\,,\qquad \si=\si(x)\,.
$$

The natural question is whether is it possible to construct 
more examples of conformal fields? Obviously, those can be 
vectors, scalars or spinors with greater number of derivatives
(see, e.g. \cite{erd,branson} for the works in this direction.).
Furthermore it can be spin-$3/2$ field with higher derivatives,
etc.
In all cases the construction of symmetric actions can be 
performed in a way described in  \cite{GBPISh97} for the 
theory (\ref{FT}). 

\section{Conformal anomaly in the semiclassical theory}

In this section we shall consider the anomalous violation
of the local conformal symmetry in the case of quantum matter 
on classical curved background. This theory is also known as
semiclassical gravity, because it shares many features with
quantum theory of gravity. The semiclassical approach is very 
important independent whether we consider it or not as an 
approximation to quantum gravity. The reason is that the 
quantum fields on curved background definitely exist in nature 
while the reality of quantum gravity is under question. It 
might occur, after all, that the gravity should not be 
quantized and, instead, it is an interaction induced by, e.g. 
(super)string theory in the low-energy limit.  

The first step is to consistently formulate the action on 
classical curved background. The standard criteria for the 
action of external metric field are (see, e.g. \cite{book})
 \ {\it a)} locality of the vacuum action, 
 \ {\it b)} renormalizability and 
 \ {\it c)} what one can call simplicity, e.g. we assume 
there are no \ $\left[m^{-1}\right]$ \ parameters or, in 
other words, we include the minimal set of terms which 
satisfy {\it a)} and  \ {\it b)} conditions.

The action of vacuum which satisfies these necessary 
conditions has the form 
\beq
S_{vac}\,=\,S_{EH}\,+\,S_{HD}\,,
\label{S_vac}
\eeq
where $\,S_{EH}\,$ is the Einstein-Hilbert action with 
cosmological term and 
\beq
S_{HD}=\int d^4x \,\sqrt{-g}\,
\left\{a_1 C^2 + a_2 E + a_3 {\Box}R + a_4 R^2 \right\}\,.
\label{S_HD}
\eeq
Here and below we use the following notations
\beq
E = R_{\mu\nu\al\be}^2 - 4 R_{\al\be}^2 + R^2\,.
\label{E}
\eeq
is the Gauss-Bonnet term (Euler density in $n=4$). We avoid
using the letter \ $G$ \ to denote this quantity because it 
may be confused with the Newton constant. 

In the case of conformal theory at the one-loop level 
it is sufficient to consider the simplified vacuum action
\beq
S_{conf.\,\,vac}\,=\,\int d^4x \sqrt{-g}
\left\{a_1C^2+a_2E+a_3{\Box}R \right\}\,.
\label{S_vac-conf}
\eeq
Let us emphasize that it is not {\it impossible} to add the 
Einstein-Hilbert action, cosmological constant or
the \ $\int\sqrt{-g}R^2$ \ term here. The statement is that 
these terms are  {\it not really necessary} at the one-loop 
level. In fact, beyond the one-loop approximation  the  
 \ $\int\sqrt{-g}R^2$ \ terms becomes also necessary, this 
means the conformal theory is not consistent beyond one loop. 
In case of broken symmetry and generated masses of the matter 
fields (e.g. through the Coleman-Weinberg mechanism), other 
mentioned terms may also become necessary. 
\vskip 2mm

Now we are in a position to consider the conformal anomaly.
We assume the theory includes the metric \ $g_{\mu\nu}$ \ as
a background field and also quantized matter fields $\Phi$.
We denote, furthermore, \ $k_\Phi$ \ the conformal weight 
of the field.

The Noether identity for the local conformal symmetry 
\beq
\left[
-\,2\,g_{\mu\nu}\,\frac{\de}{\de g_{\mu\nu}}
+\,k_\Phi\,\Phi\,\frac{\de}{\de \Phi}\right]
\,S(g_{\mu\nu},\,\Phi)\,=\,0
\eeq
produces $\,T^\mu_\mu=0\,$ on shell (\ref{loc}).

At quantum level $\,S_{vac}(g_{\mu\nu})\,$ has to be 
replaced by the effective action of vacuum 
$\,\Ga_{vac}(g_{\mu\nu})$. For the free fields only 1-loop 
order is relevant and (see \cite{book} for the introduction
and further references)
\beq
\Ga_{div}\,=\,\frac{1}{\vp}\,\int d^4x\sqrt{g}
\left\{\be_1C^2+\be_2E+\be_3{\Box}R \right\}\,,
\label{diverg}
\eeq
where \ $\vp$ \ is the regularization parameter. For instance, 
it is \ $\vp=\mu^{n-4}/(n-4)$ \ in dimensional regularization. 
In the case of global conformal symmetry, the renormalization 
group method or \ $\ze$-regularization tell us 
\cite{tmf,haw,book}
\beq
<T^\mu_\mu>\,=\,
\left\{\be_1C^2+\be_2E+a^\prime{\Box}R \right\}\,,
\label{T}
\eeq
where $a^\prime=\be_3$. However, in the case of local conformal 
invariance there is an ambiguity in the parameter \ $a^\prime$ \
\cite{birdav,duff94,AGS}. 

One can derive the anomaly in many different ways, which mainly 
differ by the regularization choice \cite{duff77,chris} (see, e.g. 
\cite{birdav} for the list of results in some regularizations).
Recently, we have analyzed the source of the ambiguity in 
full details \cite{AGS,AGBPS} and, in particular, have shown 
that the ambiguity is always related to the local terms in the 
anomaly-induced effective action of vacuum (see the next 
section) which have different form from the terms in the 
classical (conformal invariant) action. It turns out that the 
dimensional regularization does not enable one to control these
local terms and therefore the corresponding terms in the anomaly
(which are always total derivatives) remain arbitrary. On the 
other hand, in other regularizations such as point-splitting
one, there is no apparent freedom and it seems that the 
ambiguity is not there. The same is true if one derives 
the local term in the anomaly via the heat-kernel solution for 
the effective action\footnote{Let us notice that the solution 
for the anomaly-induced effective action presented in the next 
section agrees with the one obtained from the heat-lernel 
method, despite this is not easy to see \cite{Deser93}.} 
\cite{bavi3} or makes a massless limit in 
the effective action of massive fields \cite{AGS}. Finally, in 
the covariant Pauli-Villars regularization one can observe the
same ambiguity (in a somehow reduced form) and thus confirm the 
validity of the situation discovered in the dimensional 
regularization approach. We consider the mentioned ambiguity in 
some details in section 5.  

Let us consider, as an example, the derivation of anomaly 
through the most explicit method of dimensional regularization
\cite{duff77}. The theory of matter includes the following set 
of massless fields: $N_0$ scalars (spin-0), $N_{1/2}$ spinors 
(Dirac, spin-1/2) and $N_1$ Abelian vectors (spin-1). All 
$N$'s indicate a number of fields (not multiplets) 
in curved space-time, taking conformal version for scalars.
We are interested in the vacuum effects and therefore, at 
one-loop order, can restrict consideration by the free fields 
case. Using the well-known results (see, e.g.
\cite{birdav,book}) we arrive at the expression for
divergences (\ref{diverg}) with 
\beq
\be_1 &=& -\frac{1}{(4\pi)^2}\,
\Big( \frac{N_0}{120} + \frac{N_{1/2}}{20}
+ \frac{N_1}{10}\Big)\,,
\nonumber
\\
\be_2 &=& \frac{1}{(4\pi)^2}\,
\Big( \frac{N_0}{360} + \frac{11\,N_{1/2}}{360}
+ \frac{31\,N_1}{180}\Big)\,,
\nonumber
\\
\be_3 &=& -\frac{1}{(4\pi)^2}\,\Big( \frac{N_0}{180} 
+ \frac{N_{1/2}}{30} - \frac{N_1}{10}\Big)\,.
\label{divs}
\eeq

The renormalized one-loop effective action has the form
\beq
\Ga_R = S + {\bar \Ga} + \De S\,,
\label{total}
\eeq
where \ ${\bar \Ga}={\bar \Ga}_{div}+{\bar \Ga}_{fin}$ \ is 
the naive quantum correction to the classical action and 
$\De S$ is a counterterm. The classical action is
\ $S=S_{matter}+S_{vac}$, where \ $S_{vac}$ \ has the form 
(\ref{S_vac}). Indeed, only conformal invariant part of the 
vacuum action must be used in (\ref{total}).

$\De S$ in (\ref{total}) is an infinite local counterterm 
which is called to cancel the divergent part of 
${\bar \Ga}$ (\ref{divs}). Indeed $\De S$ is the only 
source of the noninvariance of the effective action, since 
naive (but divergent) contributions of quantum matter fields 
are conformal. The anomalous trace is therefore equal to
\beq
T = <T_\mu^\mu> = - \frac{2}{\sqrt{-g}}\,g_{\mu\nu}
\left.\frac{\de \,\Ga_R}{\de \,g_{\mu\nu}} \right|_{D=4}
= - \frac{2}{\sqrt{-g}}\,g_{\mu\nu}
\left.\frac{\de\,\De S}{\de\, g_{\mu\nu}} \right|_{D=4}\,.
\label{trace}
\eeq
The calculation of this expression can be done, in a most 
simple way, as follows. 
Let us change the parametrization of the metric to
\beq
{g}_{\mu\nu} = {\bar g}_{\mu\nu}\cdot e^{2\si}\,,
\label{conf}
\eeq
where $\,{\bar g}_{\mu\nu}\,$ is the fiducial metric with 
fixed determinant. There is a useful relation
\beq
 - \frac{2}{\sqrt{-g}}\,g_{\mu\nu}
\frac{\de\,A[g_{\mu\nu}]}{\de\, g_{\mu\nu}}
= - \frac{1}{\sqrt{-{\bar g}}}\,e^{- 4\si}
\left.\frac{\de\,A[{\bar g}_{\mu\nu}\,e^{2\si}]}{\de \si}
\,\right|_{{\bar g_{\mu\nu}}\rightarrow g_{\mu\nu},
\si\rightarrow 0}\,.
\label{deriv}
\eeq
At that point we need a transformation laws for the 
structures presented in (\ref{diverg}). They can be found,
for instance, in \cite{Stud}, so we will not reproduce 
these formulas here. It is sufficient to show a single 
relation between the expression depending on the original 
metric \ $g_{\mu\nu}$ \ and the transformed one  
\beq
g^\prime_{\mu\nu}=g_{\mu\nu}\,e^{2\si(x)}
\,,\qquad
\int \sqrt{-g^\prime}\,{C^\prime}^2(n)\,=\,
\int \sqrt{-g^\prime}\,e^{(n-4)\si}\,C^2(n)\,.
\label{c-trans}
\eeq
All other expressions of our interest have the same factor 
$\,e^{(n-4)\si}\,$ and, on the top of that, some extra terms 
with derivatives of $\,\si(x)$. For all terms which are not 
total derivatives, these terms are irrelevant due to the 
limit procedure in eq. (\ref{deriv}).

In the simplest case of global conformal factor 
$\si=\la=const$ we immediately arrive at the expression 
(\ref{T}) with $a^\prime=\be_3$. However in the local 
case \ $\si=\si(x)$ \ the situation is more complicated. 
It is worth mentioning that the left hand side in 
(\ref{deriv}) gives zero when applied to the integral 
of the total derivative term \ $\int\sqrt{-g}\Box R$.
We shall come back to discuss this term and the 
corresponding ambiguity in section 5. 

\section{Anomaly-induced action of vacuum}

One can use conformal anomaly to construct the equation for 
the finite part of the 1-loop correction to the effective 
action (we change notations here for the sake of convenience)
\beq
\frac{2}{\sqrt{-g}}\,g_{\mu\nu}
\frac{\de\, {\bar \Ga}_{ind}}{\de g_{\mu\nu}}
= \frac{1}{(4\pi)^2}\,
\left(\,aC^2 + bE + c{\Box} R\,\right)\,.
\label{mainequation}
\eeq
The solution of this equation is straightforward \cite{rei}
(see also generalizations for the theory with torsion 
\cite{buodsh} and with a scalar field \cite{shocom}).
The simplest possibility is to parametrize metric as in
(\ref{c-trans}), separating the conformal factor \ $\sigma(x)$ 
\ and rewrite the eq. (\ref{mainequation}) using (\ref{deriv}). 
The solution for the effective action is
$$
{\bar \Ga} = S_c[{\bar g}_{\mu\nu}] + 
\frac{1}{(4\pi)^2}\,
\int d^4 x\sqrt{-{\bar g}}\,\{ 
a\si {\bar C}^2 + b\si ({\bar E}-\frac23 {\bar {\Box}}
{\bar R}) + 2b\si{\bar \De}_4\si -
$$
\beq
- \frac{1}{12}\,(c+\frac23 b)[{\bar R} - 6({\bar \na}\si)^2 - 
({\bar \Box} \si)]^2)\}
\label{quantum}
\eeq
where $S_c[{\bar g}_{\mu\nu}]=S_c[g_{\mu\nu}]$ is an unknown 
functional of the metric, which serves as an integration 
constant for the eq. (\ref{mainequation}).

The solution (\ref{quantum}) has the merit of being simple, 
but an important disadvantage is that it is not covariant or, 
in other words, it is not expressed in terms of original 
metric $\,g_{\mu\nu}$. In order to obtain the non-local 
covariant solution and after represent it in the local 
form using auxiliary fields, we shall follow \cite{rei,a}. 

First one has to establish the following relations \cite{rei}
(see also \cite{Stud} for details):
\beq
\sqrt{-g}C^2 = \sqrt{-{\bar g}}{\bar C}^2 \,,\,\,\,\,\,\,\,\,\,
 \sqrt{-{\bar g}}\,{\bar \De}_4 = \sqrt{- {g}}\,{\De}_4\,,
\label{weyly}
\eeq
\beq
 \sqrt{-g}(E - \frac23{\Box}R) = \sqrt{-{\bar g}}({\bar E} 
- \frac23{\bar {\Box}}{\bar R} + 4{\bar {\De}}_4\si )
\label{trans Gauss}
\eeq
and also introduce the Green function for the operator 
(\ref{rei op}) \ \ $\De_{4,x}\,G(x,y)=\de(x,y)$.
Using these formulas and (\ref{deriv}) we find, for any 
$A = A({\bar g}_{\mu\nu},\si)$, the relation
\beq
\frac{\de}{\de \si (y)}\,\int d^4 x \sqrt{-g (x)}\,A\,
\left.(E 
- \frac23{\Box}R)\right|_{g_{\mu\nu} = {\bar g}_{\mu\nu}}
 = 4\sqrt{-{\bar g}}{\bar {\De}}_4 \,A = 4\sqrt{- g}{\De}_4 \,A\,.
\label{GB}
\eeq
In particular, we obtain
$$
\frac{\de}{\de \si (y)}\, \int d^4 x \sqrt{-g (x)}\, 
\int d^4 y \sqrt{-g (y)}\,\, a\,C^2(x)\,G(x,y)\,
\frac14\,\left.\Big(E - \frac23{\Box}R\Big)_y\right|_{g_{\mu\nu} 
= {\bar g}_{\mu\nu} } = 
$$
\beq
=\int d^4x\sqrt{-{\bar g} (x)}\,{\bar{\De}}_4(x)\,{\bar G}(x,y)\, 
a\,{\bar C}^2(x) = a\,\sqrt{-g}\,C^2(y)\,.
\label{wel}
\eeq
Hence the term in the effective action, which produces 
$T_a = - a C^2$, is
\beq
\Gamma_a = \frac{a}{4}\,\int d^4 x \sqrt{-g (x)}\, 
\int d^4 y \sqrt{-g (y)}\, C^2(x)
\,G(x,y)\,(E - \frac23{\Box}R)_y
\label{a-term}
\eeq
Similarly one can check that the variation (\ref{deriv}) 
produces \ $T_b = b\,(E - \frac23{\Box}R)\,$ if 
\beq
\Gamma_b = \frac{b}{8}\,\int d^4x \sqrt{-g (x)}\, 
\int d^4 y \sqrt{-g (y)}\,
(E - \frac23{\Box}R)_x\,G(x,y)\,(E - \frac23{\Box}R)_y
\label{b-term}
\eeq

Finally, the third constituent of the induced action is 
the local expression
\beq
\Ga_{c} = - \frac{c+\frac23\,b}{12(4\pi)^2}
\,\int d^4 x \sqrt{-g (x)}\,R^2(x) \,.
\label{c-term}
\eeq
The covariant solution of eq. (\ref{mainequation})
is a sum of (\ref{a-term}),(\ref{b-term}) and (\ref{c-term}).

Our next task is to rewrite the nonlocal expressions 
obtained above using some auxiliary scalar fields. 
Let us notice that there are two distinct ways of doing 
that, leading to the slightly different results. The first 
option is to introduce the auxiliary fields as a quantum 
objects, such that, after Gaussian integration over them, 
we should come back to the non-local expression described 
above. Another possibility \footnote{Author is grateful to
Roberto Balbinot for discussion concerning this point.}
is to consider auxiliary fields as purely classical 
objects. After 
using the classical equations of motion for these fields
and replacing them back to the action we come to the 
original non-local expressions. The difference between 
the two approaches is that, in the first case, one has
to account for the contributions of the auxiliary fields 
to the anomaly. As a result, the coefficients $a,b,c$ in
(\ref{mainequation}) get modified \cite{rei}. In this 
article we will not account for this modification and 
follow the second approach.  

As a first step the remaining terms can be rewritten in 
the symmetric form
\beq
\Gamma_{a,b} 
&=& \int d^4 x \sqrt{-g (x)} \int d^4 y \sqrt{-g (y)}
(E - \frac23{\Box}R)_x G(x,y)\left[\frac{a}{4}C^2 
- \frac{b}{8}(E - \frac23{\Box}R)\right]_y 
\nonumber
\\
&=& -\frac{b}{8}\int d^4 x d^4 y \sqrt{g (y)g (x)}
\left[(E - \frac23{\Box}R) - 
\frac{a}{b}C^2\,\right]_x 
G(x,y)\left[(E - \frac23{\Box}R) - \frac{a}{b}C^2\right]_y 
\nonumber
\\
&+& \frac12\,\int d^4 x d^4 y \sqrt{g (y)g (x)}\,
\left(\,\frac{a}{2\sqrt{b}}\,C^2\,\right)_x\,G(x,y)\,
\left(\,\frac{a}{2\sqrt{b}}\,C^2\,\right)_y
\label{nonloc}
\eeq
The last two terms are appropriate objects for rewriting them 
using the auxiliary fields. In this way we arrive at the following 
final expression for the anomaly generated effective action of 
gravity.
\beq
\Gamma &=& S_c[g_{\mu\nu}] 
- \frac{3c+2b}{36(4\pi)^2}\,\int d^4 x \sqrt{-g (x)}\,R^2(x) 
+  \int d^4 x \sqrt{-g (x)}\,\Big\{
\frac12 \,\ph\De_4\ph - \frac12 \,\psi\De_ 4\psi
\nonumber
\\
&+& \ph\,\left[\,\frac{\sqrt{b}}{8\pi}\,(E -\frac23\,{\Box}R)\,
- \frac{a}{8\pi\sqrt{b}}\,C^2\,\right] 
+ \frac{a}{8\pi \sqrt{b}}\,\psi\,C^2 \,\Big\}\,.
\label{finaction}
\eeq

Some remarks are in order.

1) The local covariant form (\ref{finaction}) is dynamically 
equivalent to the non-local covariant form. The complete 
definition of the Cauchy problem in the theory with the 
non-local action requires defining the boundary conditions for 
the Green functions \ $G(x,y)$, which shows up independently 
in the two terms \ (\ref{a-term}) \ and \ (\ref{b-term}). The 
same can be achieved, in the local version, by imposing the 
boundary conditions on the two auxiliary fields $\ph$ and $\psi$.

2) The kinetic term for the auxiliary field $\ph$ is positive 
while for $\si$ it was negative. For $\psi$ the kinetic term 
has negative sign. The wrong sign does not lead to problems 
here, because both fields are auxiliary and do not propagate
independently. 

3) We introduced the new structure 
\ $\int C^2_x G(x,y) C^2_y$ \ into the action, despite it 
was not directly produced by anomaly. This term is indeed 
conformal invariant and therefore its emergence may be viewed 
as a simple redefinition of the conformal invariant functional 
$S_c[g_{\mu\nu}]$. On the other hand, writing the non-conformal 
terms in the symmetric form (\ref{nonloc}), we have modified 
the four point function in a very essential way. Therefore, 
introducing the mentioned conformal term we have just restored 
the basic structure of the terms generated by anomaly. For this
reason, the emergence of the second auxiliary scalar \cite{a} 
is not an artificial procedure but represents a necessary 
element of writing the induced action in a local 
form\footnote{The effective action (ref{finaction}) has 
been introduced in the paper \cite{a}. Qualitatively similar
manner of introducing second scalar has been suggested later 
on in \cite{MaMo}.}. 

4) The second scalar also proved useful for applications. 
In particular, the vacuum states of the black hole (Boulware, 
Hartle-Hawking and Unruh) can be classified through the 
choice of initial conditions for the two auxiliary fields 
\cite{balsan}. Let us stress that this can not be 
accomplished by using only one field $\ph$. Therefore the 
correspondence with other approaches to Hawking radiation 
indicates that our considerations about the correctness 
of introducing the second auxiliary scalar are correct. 

5) Another important application of the anomaly-induced
action is the model of anomaly-induced inflation 
\cite{anju,shocom}, or Modified Starobinsky Model. In 
this case the behaviour of conformal factor of the metric
is not affected by the presence of the second auxiliary 
scalar. However, for investigating the evolution of 
gravitational waves specifying the initial data for both 
scalars is essential and the situation is 
close to the one in the black hole case. 

\section{Ambiguity of local/surface anomalous terms}

The ambiguity of the local anomalous term \ $\int\sqrt{-g}R^2$
in the effective action and the corresponding term \ $\Box R$
in the anomaly can be observed either in dimensional or 
in covariant Pauli-Villars regularization \cite{AGS}.

Let us start from the dimensional regularization and come back 
to the relation (\ref{trace}). As we have already mentioned 
in section 3, the counterterm  \ $\int\sqrt{-g}\Box R$ \ does
not contribute to the anomaly of local conformal symmetry. 
Hence the anomaly comes from the  \ $\int\sqrt{-g}C^2(d)$-type
counterterm. However, the requirements of finiteness of 
renormalized effective and the locality leave us the freedom 
in choosing the parameter $d$. If we take \ $d=n+\ga\cdot[n-4]$,
where \ $\ga$ \ is an arbitrary parameter, we meet 
\ $a^\prime \sim \ga$ \ and therefore the coefficient \ $a$
is arbitrary. It is easy to see that this arbitrariness 
is equivalent to adding  \ $\int R^2$-term to the classical 
action.

The same result can be achieved in the covariant Pauli-Villars 
regularization, where one has to introduce a set of massive 
``regulator'' fields. For example, in the case of a massless
conformal scalar \ $\ph$ \ we have to start from the action 
\beq
S_{\rm reg} = \sum_{i=0}^N \int d^4 x \sqrt{g}
\left\{
(\na \ph_i)^2 + (\xi_i R + m^2_i) \ph_i^2 \right\}.
\label{PV}
\eeq
The physical scalar field $\varphi \equiv\varphi_0$ 
is conformal $\xi={1}/{6}$, $m_0=0$ and bosonic $s_0=1$,
while PV regulators $\varphi_i$ are massive $m_i=\mu_i M$ 
and can be bosonic $s_i=1$ or fermionic $s_i=-2$. 

The UV limit $M \to \infty$ produces the vacuum
Eff. Action. The calculation is based on our result for 
the EA of the massive scalar. We assume that the Pauli-Villars 
regulators may have conformal \ $\xi_i = 1/6$ or non-conformal 
couplings \ $\xi_i\neq 1/6$. 

The regularized effective action is
\beq
{\bar \Ga}^{(1)}_{\rm reg}\,=\, 
\lim_{\Lambda\to \infty}\sum_{i=0}^N s_i 
{\bar \Ga}^{(1)}_{\rm i} 
\left(m_i,\xi_i,\Lambda\right)\,, 
\label{regEA}
\eeq
where \ $\Lambda$ \ is an auxiliary momentum cut-off. It is 
important that we possess the explicit expression of the 
\ ${\cal O}(R^2)$ \ effective action of massive fields 
derived in \cite{apco}. The conformal anomaly is \cite{AGS}
\beq
T= \frac{1}{(4\pi)^2}\Big[\frac{1}{180} E - \frac{1}{120}C^2  
+ \Big(12\de - \frac{1}{180}\Big){\Box}R\Big]\,, \quad
\delta=\sum_{i=1}^N s_i\Big(\xi_i-\frac16\Big)^2\ln\mu_i^2\,, 
\label{regEA-1}
\eeq
Exactly as in the dimensional regularization case, the 
ambiguity is equivalent to the freedom of adding the finite 
 $\int\sqrt{-g}R^2$ term and can be fixed only by imposing 
the renormalization condition for this term. 
The qualitative result is that the 
definition of the local finite part of the quantum 
corrections is ambiguous, even if the corresponding term 
is not present in the classical action and is not 
renormalized. In order to fix the ambiguity one has to 
use renormalization condition and for that it is necessary 
to introduce the non-conformal local term into the classical 
action. As a result the theory is not conformal anymore. 
This consideration can be generalized for the case of 
more general (non purely metric) backgrounds \cite{AGBPS},
where the situation is similar albeit somehow more 
complicated. 

\section{Weyl quantum gravity}

Finally, let us consider the problem of anomalous violation 
of local conformal symmetry in the conformal Weyl quantum 
gravity. The action of the theory has the form
\beq
S_W= \int d^4x\sqrt{g}
\left\{\frac{1}{2\la}C^2-\frac{1}{\rho}E+\tau\Box R\right\}\,,
\label{Weyl-1}
\eeq
The action
(\ref{Weyl-1}) is conformal invariant in a sense it satisfies 
the conformal Noether identity 
\beq 
-\,\,\frac{2}{\sqrt{-g}}\,
g_{\mu\nu}\, \frac{\de S_W}{\de g_{\mu\nu}}\,=\,0 \,.
\label{Noether} 
\eeq
The conformal theory (\ref{Weyl-1}) may be an interesting 
model of quantum gravity \cite{frts82,bush85,book},
while General Relativity may be induced at quantum level 
if the Weyl gravity is coupled to a conformal scalar 
\cite{Buch-Weyl,induce,brv}. 

The theory (\ref{Weyl-1}) is the particular case of the 
general higher derivative quantum gravity which is 
know to be renormalizable \cite{stelle,vortyu}. At the 
same time the properties of conformal theory may be 
quite different from the general one. 
One can formulate two main questions concerning the 
properties of the conformal theory at quantum level. 

1) Whether the one-loop and/or higher loop infinite 
renormalization of this theory is conformal invariant.
In other words, whether the conformal theory is 
multiplicatively renormalizable. 

2) Whether the anomalous violation of local conformal 
symmetry occurs in the finite part of the effective 
action. If so, we need to know whether the corresponding 
ambiguities, similar to the ones discussed in the previous
section, take place here. 

The derivation of the one-loop divergences in the conformal 
theory has been performed in \cite{frts82,antmot,Weyl}. 
The result obtained in \cite{Weyl} with the method including 
rigid automatic control of the calculations fits with the
previous ones and has the form 
\beq
\Ga ^{(1)}_{{\rm div}} = -
\frac{1}{\ep}\int d^4 x\sqrt{-g}
\left\{\frac{87}{20} E - \frac{199}{30}C^2\right\}\,.
\label{1-loop}
\eeq
In $n=4$, the dependence on the Gauss-Bonnet term is 
absent, as it was anticipated earlier \cite{frts82,antmot}. 
At the same time this dependence is essential for the 
renormalization group near four dimensions, producing 
a number of new nontrivial fixed points, some of them 
UV stable and some IR stable.  
There is no real need to calculate the remaining 
\ $\int\Box R$-type counterterm, because it is gauge
fixing dependent \cite{frts82,a}.
According to (\ref{1-loop}), there is no need for 
the \ $\int d^4x\sqrt{-g}R^2$-type  counterterm and, 
correspondingly, no need to use the so-called
special conformal regularization \cite{truffin,frts82}.
At one loop the theory is multiplicatively renormalizable
in the usual sense. 

The derivation of anomaly may proceed in exactly the 
same way as in the semiclassical theory. In particular, 
the anomaly associated to the divergences (\ref{1-loop})
is well defined and the corresponding non-local terms 
in the induced action can be obtained in a unique way. 
At the same time, the local \ $\int\sqrt{-g}R^2$-term
is plagued by double ambiguity: due to the gauge 
dependence of the corresponding \ $\Box R$-type 
divergence and because of the renormalization-scheme
dependence. Finally, the only way to arrive at the 
well defined quantum corrections is to violate the 
conformal symmetry at the classical level already. 
In case this violation is weak, the quantum corrections
will respect the corresponding hierarchy. 

\section{Conclusions}

We have reviewed some important aspects of local conformal
symmetry and in particular its violation at the quantum 
level by anomaly. There is a variety of theories with local
conformal symmetry in $n=4$. Along with the conventional 
scalar, spinor and vector cases, there are different 
higher derivative conformal models with higher derivatives. 

In the semiclassical theory local conformal symmetry is 
violated by the trace anomaly. There are both local and 
non-local terms in the effective action behind the anomaly,
but the local terms are plagued by ambiguities, indicating 
certain inconsistency in the quantum corrections. 
For conformal quantum gravity similar ambiguities produce
total inconsistency of the theory beyond the one loop level,
because at higher loops the emergence of the non-symmetric
counterterms looks unavoidable. So, we can conclude that,
in general, conformal invariant theories are not consistent 
at quantum level. In fact, the local conformal symmetry may 
be only approximate, despite it is a very useful tool for 
calculating quantum corrections.  
\vskip 6mm


{\large\bf Acknowledgments.} 

Author is very grateful to his collaborators on the subject, 
especially to M. Asorey, E. Gorbar, G. de Berredo-Peixoto
and J. Sol\`a 
for numerous discussions and common works. I am also 
thankful to the organizers of the Meeting for invitation 
to give this talk and for support. The work has been partially 
supported by the PRONEX project and research grants from 
FAPEMIG (MG, Brazil) and CNPq (Brazil) and by the fellowships 
from CNPq and ICTP (Italy). The text 
of this Proceedings has been completed during my visit to 
Theory Division of CERN and I would like to thank them 
for support and hospitality.


\begin{thebibliography}{99}

\bibitem{faraoni} V. Faraoni, E. Gunzig and P. Nardone,
{\it Conformal transformations in classical gravitational 
theories and in cosmology,}
Fund. Cosmic Phys. {\bf 20} (1999) 121 [gr-qc/9811047]. 

\bibitem{anju} J.C.Fabris, A.M.Pelinson and I.L.Shapiro,
{\it On the gravitational waves on the background
of anomaly-induced inflation,}
Nucl. Phys. {\bf B597} (2001) 539 [hep-th/0009197]. 

\bibitem{conf} I.L. Shapiro and H. Takata, 
{\it One-loop renormalization of the four-dimensional theory for 
quantum dilaton gravity,}
Phys. Rev. {\bf D52} (1995) 2162 [hep-th 9502111];
{\it Conformal transformation in gravity,}
Phys. Lett. {\bf B361} (1995) 31 [hep-th/9504162].

\bibitem{ohanlon} J. O'Hanlon, 
{\it Intermediate-range gravity - a generally covariant model,}
Phys. Rev. Lett. {\bf 29} (1972) 137.

\bibitem{Adler} S.L. Adler, 
{\it Einstein gravity as a symmetry breaking effect 
in quantum field theory,} Rev. Mod. Phys. {\bf 54} (1982) 729.
 
\bibitem{Buch-Weyl} I. L. Buchbinder, 
{\it Mechanism For Induction Of Einstein Gravitation,}
Sov. J. Phys. {\bf 31} (1986) 77.

\bibitem{induce} I.L. Shapiro, {\it Hilbert-Einstein 
Action from Induced Gravity coupled with Scalar Field,}
Mod. Phys. Lett. {\bf 9A} (1994) 1985 [hep-th/9403077].

\bibitem{brv} I.L. Shapiro and G. Cognola,
{\it Interaction of Low - Energy Induced Gravity with 
Quantized Matter and Phase Transition Induced by Curvature,}
Phys. Rev. {\bf D51} (1995) 2775 [hep-th/9406027]; 
{\it Back reaction of vacuum and the renormalization group 
flow from the conformal fixed point,}
Class. Quant. Grav. {\bf 15} (1998) 3411 [hep-th/9804119].

\bibitem{Paneitz}
S. Paneitz, {\it A Quartic Conformally Covariant Differential
Operator for Arbitrary Pseudo-Riemannian Manifolds,} 
MIT preprint, 1983.

\bibitem{rei} R.J. Riegert, 
{\it A Nonlocal Action For The Trace Anomaly.}
Phys. Lett. {\bf B134} (1984) 56;

E.S. Fradkin and A.A. Tseytlin, 
{\it Conformal Anomaly In Weyl Theory And Anomaly Free 
Superconformal Theories,} Phys. Lett. {\bf B134} (1984) 187.

\bibitem{ai} I. Antoniadis, J. Iliopoulos and T. N. Tomaras,
{\it On The Stability Of Background Solutions In Conformal Gravity,} 
Nucl. Phys. {\bf B261} (1985) 157.

\bibitem{acacio} J. A. de Barros and I.L. Shapiro, 
{\it Renormalization group study of the higher derivative 
conformal scalar model,} 
Phys. Lett. {\bf B412} (1997) 242 [hep-th/9706123].

\bibitem{eli} 
E. Elizalde, A.G. Zhecksenaev, S.D. Odintsov and I.L. Shapiro,
{\it One-loop renormalization and asymptotic behavior of a
higher-derivative scalar theory in curved spacetime,}
Phys. Lett.  {\bf B328} (1994) 297 [hep-th/9402154].

\bibitem{FrTs85} E.S. Fradkin and A.A. Tseytlin, 
{\it Conformal Supergravity,}
Phys. Repts. {\bf 119} (1985) 233.

\bibitem{GBPISh97} G. de Berredo-Peixoto and I.L. Shapiro,
{\it The higher derivative fermionic operator and trace anomaly,}
Phys. Lett. {\bf B514} (2001) 377 [hep-th/0101158].

\bibitem{erd} J. Erdmenger, 
{\it Conformally covariant differential operators: 
Properties and applications,}
Class. Quant. Grav. {\bf 14} (1997) 2061 [hep-th/9704108];
{\it Conformally covariant differential operators: 
Symmetric tensor fields.}
J. Erdmenger and H. Osborn, Class. Quant. Grav. 
{\bf 15} (1998) 273 [gr-qc/9708040].

\bibitem{branson}  T.P. Branson,
{\it Conformaly covariant equations on differential forms,}
Comm. Part. Diff. Eq. {\bf 7} (1982) 393; 
{\it An anomaly associated with 4-dimensional quantum gravity,}
Commun. Math. Phys. {\bf 178} (1996) 301.
 
\bibitem{book} I.L. Buchbinder, S.D. Odintsov, I.L. Shapiro,
{\it Effective Action in Quantum Gravity}, IOP Publishing,
Bristol, 1992.

\bibitem{tmf} I.L. Buchbinder, 
{\it Renormalization-group equations in curved space-time,}
Theor. Math. Phys. {\bf 61} (1984) 393.

\bibitem{haw} S.W. Hawking, 
{\it Zeta Function Regularization Of Path Integrals In 
Curved Space-Time,} Comm. Math. Phys. {\bf 55} (1977) 133.

\bibitem{birdav} N.D. Birrell, P.C.W. Davies,
{\it Quantum fields in curved space}, 
Cambridge Univ. Press, Cambridge, 1982.

\bibitem{duff94} M.J. Duff,  
{\it Twenty years of the Weyl anomaly,}
Class. Quant. Grav. {\bf 11} (1994) 1387 [hep-th/9308075]. 

\bibitem{AGS} M. Asorey, E.V. Gorbar and I.L. Shapiro,
{\it Universality and Ambiguities of the Conformal Anomaly.}
Class. Quant. Grav. {\bf 21} (2003) 163 [hep-th/0307187]. 

\bibitem{duff77} M.J. Duff, 
{\it Observations On Conformal Anomalies,}
Nucl. Phys. {\bf B125} (1977) 334;

S. Deser, M.J. Duff and C. Isham, 
{\it Nonlocal Conformal Anomalies,}
Nucl. Phys. {\bf B111} (1976) 45.

\bibitem{chris} S.M. Christensen, 
{\it Vacuum Expectation Value Of The Stress Tensor In An Arbitrary 
Curved Background: The Covariant Point Separation Method,}
Phys. Rev. {\bf D14} (1976) 2490; 
{\it Regularization, Renormalization, And Covariant 
Geodesic Point Separation,} {\bf D17} (1978) 948.

\bibitem{AGBPS}
M. Asorey, G. de Berredo-Peixoto and I.L. Shapiro,
{\it Renormalization Ambiguities and Conformal Anomaly
in  Metric-Scalar Backgrounds}, hep-th/0609138. 

\bibitem{Deser93} S. Deser and A. Schwimmer,
{\it Geometric Classification of Conformal Anomalies in 
Arbitrary Dimensions,}
Phys. Lett. {\bf 309B} (1993) 279 [hep-th/9302047]
 
S. Deser, 
{\it Closed form effective conformal anomaly actions 
in $D \geq 4$,}
Phys. Lett. {\bf 479B} (2000) 315 [hep-th/9911129]. 

\bibitem{bavi3} A.O. Barvinsky, Yu.V. Gusev, G.A. Vilkovisky 
and V.V. Zhitnikov, 
{\it The One loop effective action and trace anomaly in 
four-dimensions,} Nucl.Phys. {\bf B439} (1995) 561 [hep-th/9404187]. 
 
\bibitem{Stud}
D. F. Carneiro, E. A. Freitas, B. Gon\c{c}alves, 
A. G. de Lima and I. L. Shapiro,
{\it On Useful Conformal Tranformations In General Relativity,}
Grav. and Cosm. {\bf 40} (2004) 305 [gr-qc/0412113].

\bibitem{buodsh}
I.L. Buchbinder, S.D. Odintsov and I.L. Shapiro,
{\it Nonsingular cosmological model with torsion induced by 
vacuum quantum effects,} Phys. Lett. {\bf B162} (1985) 92;

J.A. Helayel-Neto, A. Penna-Firme and I. L. Shapiro,
{\it Conformal symmetry, anomaly and effective action 
for metric-scalar gravity with torsion,} 
Phys. Lett. {\bf B479} (2000) 411 [gr-qc/9907081]. 

\bibitem{shocom} I.L. Shapiro, J. Sol\`{a}, 
{\it Massive fields temper anomaly-induced inflation,}
Phys. Lett. {\bf B530} (2002) 10 [hep-ph/0104182];

A.M. Pelinson, I.L. Shapiro, F.I. Takakura,
{\it On the stability of the anomaly-induced inflation,}
Nucl. Phys. {\bf B648} (2003) 417 [hep-ph/0208184].

\bibitem{a} I. L. Shapiro and A. G. Jacksenaev,
{\it Gauge dependence in higher derivative quantum gravity 
and the conformal anomaly problem,} 
Phys. Lett. {\bf B324}  (1994) 284.

\bibitem{MaMo} P. O. Mazur and E. Mottola, 
{\it Weyl cohomology and the effective action for conformal 
anomalies}, Phys. Rev. {\bf D64} (2001) 104022. 

\bibitem{balsan} R. Balbinot, A. Fabbri and I.L. Shapiro,
{\it Anomaly induced effective actions and Hawking radiation,}
Phys. Rev. Lett. 83 (1999) 1494 [hep-th/9904074]; \
{\it Vacuum polarization in Schwarzschild space-time
by anomaly induced effective actions and Hawking radiation,}
Nucl. Phys.  {\bf B559} (1999) 301 [hep-th/9904162].

\bibitem{apco} E.V. Gorbar, I.L. Shapiro,
{\it Renormalization Group and Decoupling in Curved Space.}
JHEP {\bf 02} (2003) 021 [hep-ph/0210388];
{\it Renormalization Group and Decoupling in Curved Space:
II. The Standard Model and Beyond.}
JHEP {\bf 06} (2003) 004 [hep-ph/0303124].

\bibitem{frts82} E.S. Fradkin and  A.A. Tseytlin,
{\it Higher Derivative Quantum Gravity: One Loop 
Counterterms And Asymptotic Freedom}, 
Nucl. Phys. {\bf B201} (1982) 469.

\bibitem{bush85} I.L. Buchbinder and  I.L. Shapiro,
{\it On the influence of the gravitational interaction on the
behavior of the effective constants of Yukawa and scalar 
coupling,} Sov. J. Nucl. Phys. {\bf 44} (1986) 1033.

\bibitem{stelle} K.S. Stelle,  
{\it Renormalization Of Higher Derivative Quantum Gravity}, 
Phys. Rev. {\bf D16} (1977) 953.

\bibitem{vortyu} B.L. Voronov and I.V. Tyutin,
{\it On renormalization of $R^2$ gravitation,}
Yad. Fiz. (Sov. Journ. Nucl. Phys.) {\bf 39} (1984) 998.

\bibitem{antmot} I. Antoniadis, P.O. Mazur and E. Mottola,
{\it Conformal symmetry and central charges in four-dimensions,}
Nucl. Phys. {\bf B388} (1992) 627 [hep-th/9205015].

\bibitem{Weyl} G. de Berredo-Peixoto and I.L. Shapiro,
{\it Conformal Quantum Gravity with the Gauss-Bonnet term},
Phys. Rev. {\bf D70} (2004) 044024 [hep-th/0307030].

\bibitem{truffin} F. Englert, C. Truffin and R. Gastmans,
{\it Conformal Invariance In Quantum Gravity,}
Nucl.Phys. {\bf B117} (1976) 407;

E.S. Fradkin and G.A. Vilkovisky, 
{\it Conformal Off Mass Shell Extension And Elimination Of 
Conformal Anomalies In Quantum Gravity,}
Phys. Lett. {\bf B73} (1978) 209.

\end{thebibliography}
\end{document}